\documentstyle[aps,prl,epsfig,multicol]{revtex}      

\newcommand{\pythjet}{\textsc{Py\-thia/Jet\-set}}
\newcommand{\cerenkov}{\v{C}erenkov}
\newcommand{\Kstar}{\overline{K}^{\;\!*0}}
\newcommand{\eg}{e.g.}

\newcommand{\etal}{{\em et al.}}

\def\issue(#1,#2,#3){{\bf #1}, #2 (#3)} 

\def\opcit(#1){ {\em op. cit.}, #1}

\def\APP(#1,#2,#3){Acta Phys.\ Polon.\ \issue(#1,#2,#3)}
\def\ARNPS(#1,#2,#3){Ann.\ Rev.\ Nucl.\ Part.\ Sci.\ \issue(#1,#2,#3)}
\def\CPC(#1,#2,#3){Comp.\ Phys.\ Comm.\ \issue(#1,#2,#3)}
\def\CIP(#1,#2,#3){Comput.\ Phys.\ \issue(#1,#2,#3)}
\def\CJP(#1,#2,#3){Chin.\ J.\ Phys. (Taipei)\ \issue(#1,#2,#3)}
\def\EPJC(#1,#2,#3){Eur.\ Phys.\ J.\ C\ \issue(#1,#2,#3)}
\def\EPJD(#1,#2,#3){Eur.\ Phys.\ J.\ direct\ C\ \issue(#1,#2,#3)}
\def\IEEETNS(#1,#2,#3){IEEE Trans.\ Nucl.\ Sci.\ \issue(#1,#2,#3)}
\def\MPL(#1,#2,#3){Mod.\ Phys.\ Lett.\ \issue(#1,#2,#3)}
\def\NP(#1,#2,#3){Nucl.\ Phys.\ \issue(#1,#2,#3)}
\def\NIM(#1,#2,#3){Nucl.\ Instrum.\ Meth.\ \issue(#1,#2,#3)}
\def\PL(#1,#2,#3){Phys.\ Lett.\ \issue(#1,#2,#3)}
\def\PRD(#1,#2,#3){Phys.\ Rev.\ D \issue(#1,#2,#3)}
\def\PRL(#1,#2,#3){Phys.\ Rev.\ Lett.\ \issue(#1,#2,#3)}
\def\SJNP(#1,#2,#3){Sov.\ J. Nucl.\ Phys.\ \issue(#1,#2,#3)}
\def\ZPC(#1,#2,#3){Zeit.\ Phys.\ C \issue(#1,#2,#3)}

\begin{document}

\draft 		

\title{
Search for Rare and Forbidden Charm Meson Decays 
$D^0\rightarrow V\ell^+\ell^-$ and $hh\ell \ell$ 
}

\author{
    E.~M.~Aitala,$^9$
       S.~Amato,$^1$
    J.~C.~Anjos,$^1$
    J.~A.~Appel,$^5$
       D.~Ashery,$^{14}$
       S.~Banerjee,$^5$
       I.~Bediaga,$^1$
       G.~Blaylock,$^8$
    S.~B.~Bracker,$^{15}$
    P.~R.~Burchat,$^{13}$
    R.~A.~Burnstein,$^6$
       T.~Carter,$^5$
    H.~S.~Carvalho,$^{1}$
    N.~K.~Copty,$^{12}$
    L.~M.~Cremaldi,$^9$
       C.~Darling,$^{18}$
       K.~Denisenko,$^5$
       S.~Devmal,$^3$
       A.~Fernandez,$^{11}$
    G.~F.~Fox,$^{12}$
       P.~Gagnon,$^2$
       C.~Gobel,$^1$
       K.~Gounder,$^9$
    A.~M.~Halling,$^5$
       G.~Herrera,$^4$
       G.~Hurvits,$^{14}$
       C.~James,$^5$
    P.~A.~Kasper,$^6$
       S.~Kwan,$^5$
    D.~C.~Langs,$^{12}$
       J.~Leslie,$^2$
       B.~Lundberg,$^5$
       J.~Magnin,$^1$
       S.~MayTal-Beck,$^{14}$
       B.~Meadows,$^3$
 J.~R.~T.~de~Mello~Neto,$^1$
       D.~Mihalcea,$^7$
    R.~H.~Milburn,$^{16}$
    J.~M.~de~Miranda,$^1$
       A.~Napier,$^{16}$
       A.~Nguyen,$^7$
    A.~B.~d'Oliveira,$^{3,11}$
       K.~O'Shaughnessy,$^2$
    K.~C.~Peng,$^6$
    L.~P.~Perera,$^3$
    M.~V.~Purohit,$^{12}$
       B.~Quinn,$^9$
       S.~Radeztsky,$^{17}$
       A.~Rafatian,$^9$
    N.~W.~Reay,$^7$
    J.~J.~Reidy,$^9$
    A.~C.~dos Reis,$^1$
    H.~A.~Rubin,$^6$
    D.~A.~Sanders,$^9$
 A.~K.~S.~Santha,$^3$
 A.~F.~S.~Santoro,$^1$
       A.~J.~Schwartz,$^{3}$
       M.~Sheaff,$^{17}$
    R.~A.~Sidwell,$^7$
    A.~J.~Slaughter,$^{18}$
    M.~D.~Sokoloff,$^3$
       J.~Solano,$^1$
    N.~R.~Stanton,$^7$
    R.~J.~Stefanski,$^5$  
       K.~Stenson,$^{17}$ 
    D.~J.~Summers,$^9$
       S.~Takach,$^{18}$
       K.~Thorne,$^5$
    A.~K.~Tripathi,$^{7}$
       S.~Watanabe,$^{17}$
 R.~Weiss-Babai,$^{14}$
       J.~Wiener,$^{10}$
       N.~Witchey,$^7$
       E.~Wolin,$^{18}$
    S.~M.~Yang,$^7$
       D.~Yi,$^9$
       S.~Yoshida,$^7$
       R.~Zaliznyak,$^{13}$ and
       C.~Zhang$^7$ \\
\begin{center} (Fermilab E791 Collaboration) \end{center}
}

\address{
\begin{center}
\tabcolsep=0.5pt
\begin{tabular}{rlcrl}
$^1$ & Centro Brasileiro de Pesquisas F\'\i sicas, Rio de Janeiro, Brazil & &
$^2$ & University of California, Santa Cruz, CA 95064\\
$^3$ & University of Cincinnati, Cincinnati, OH 45221 & &
$^4$ & CINVESTAV, 07000 Mexico City, DF Mexico\\
$^5$ & Fermilab, Batavia, IL 60510 & &
$^6$ & Illinois Institute of Technology, Chicago, IL 60616\\
$^7$ & Kansas State University, Manhattan, KS 66506 & &
$^8$ & University of Massachusetts, Amherst, MA 01003\\
$^9$ & University of Mississippi--Oxford, University, MS 38677 & &
$^{10}$ & Princeton University, Princeton, NJ 08544\\
$^{11}$ & Universidad Autonoma de Puebla, Puebla, Mexico & &
$^{12}$ & University of South Carolina, Columbia, SC 29208\\
$^{13}$ & Stanford University, Stanford, CA 94305 & &
$^{14}$ & Tel Aviv University, Tel Aviv, Israel\\
$^{15}$ & Box 1290, Enderby, British Columbia V0E 1V0, Canada & &
$^{16}$ & Tufts University, Medford, MA 02155\\
$^{17}$ & University of Wisconsin, Madison, WI 53706 & \hspace*{4mm} &
$^{18}$ & Yale University, New Haven, CT 06511\\
\end{tabular}
\end{center}
}

\date{\today}

\maketitle

\begin{abstract}
We report results of a search for flavor-changing neutral current 
(FCNC), lepton-flavor, and lepton-number violating decays of the 
$D^0$ (and its antiparticle) into 3 and 4-bodies. Using data from 
Fermilab charm hadroproduction experiment E791, we examine modes 
with two leptons (muons or electrons) and either a $\rho ^0$, 
$\Kstar$, or $\phi$ vector meson or a non-resonant $\pi \pi$, 
$K\pi $, or $KK$ pair of pseudoscalar mesons. No evidence for any 
of these decays is found. Therefore, we present branching-fraction 
upper limits at 90$\%$ confidence level for the 27 decay modes 
examined (18 new).
\end{abstract}


\begin{multicols}{2}
The E791 Collaboration has previously reported limits on rare 
and forbidden dilepton decays of the charged charm $D$ meson 
\cite{FCNC,FCNCnew}. Such measurements probe the SU(2)$\times $U(1) 
Standard Model of electroweak interactions in search of new mediators 
and couplings \cite{Pakvasa,SCHWARTZ93}. We extend the methodology to 
27 dilepton decay modes of the neutral $D$ meson. The modes are 
resonant $D^0\rightarrow V\ell^+\ell^-$ decays, where $V$ is a 
$\rho ^0$, $\Kstar$, or $\phi$, and non-resonant 
$D^0\rightarrow hh\ell \ell $ decays, where $h$ is a $\pi$ or $K$. The 
leptons are either muons or electrons. Charge-conjugate modes are 
implied. The modes are lepton flavor-violating 
(\eg, \ $D^{0}\rightarrow \rho ^{0}\mu ^{+}e^{-}$), or 
lepton number-violating 
(\eg, \ $D^{0}\rightarrow \pi ^{-}\pi ^{-}\mu ^{+}\mu ^{+}$), or 
flavor-changing neutral current decays 
(\eg, \ $D^{0}\rightarrow \Kstar e^{+}e^{-}$). Box diagrams can mimic 
FCNC decays, but only at the $10^{-10}$ to $10^{-9}$ 
level \cite{SCHWARTZ93,Fajfer}. Long range effects 
(\eg, \ $D^{0}\rightarrow \Kstar \rho ^{0}, \ 
\rho^{0}\rightarrow  e^{+}e^{-}$) can occur at the $10^{-6}$ 
level \cite{Fajfer,Singer}. Numerous experiments have studied rare 
decays of charge -1/3 strange quarks. Charge 2/3 charm quarks are 
interesting because they might couple differently \cite{Castro}.

The data come from measurements with the Fermilab E791 
spectrometer\cite{e791spect}. A total of $2 \times 10^{10}$ events were 
taken with a loose transverse energy requirement. These events were 
produced by a 500 GeV/$c$~ $\pi ^{-}$ beam 
interacting in a fixed target consisting of five thin, well-separated 
foils. Track and vertex information came from ``hits'' in 23 silicon 
microstrip planes and 45 wire chamber planes. This information and the 
bending provided by two dipole magnets were used for momentum analysis 
of charged particles. Kaon 
identification was carried out by two multi-cell \cerenkov{} counters 
that provided $\pi /K$ separation in the momentum range $6-60$~GeV/$c$ 
\cite{Bartlett}. We required that the momentum-dependent light yield in 
the \cerenkov{} counters be consistent for kaon-candidate tracks, 
except for those in decays with $\phi \rightarrow  K^{+}K^{-}$, where 
the narrow mass window for the $\phi$ decay provided sufficient kaon 
identification (ID). 

Electron ID was based on transverse shower shape plus matching wire 
chamber tracks to shower positions and energies in 
an electromagnetic calorimeter \cite{SLIC}.
The electron ID efficiency varied from 62$\%$ below 9 GeV/$c$ to 45$\%$ 
above 20 GeV/$c$. The probability to misidentify a pion as an electron 
was $\sim 0.8\%$, independent of pion momentum.

Muon ID was obtained from two planes of scintillation 
counters. The first plane (5.5 m $\times$ 3.0 m) of 15 counters 
measured the horizontal position while the second plane (3.0 m $\times$ 
2.2 m) of 16 counters measured the vertical position. There were 
about 15 interaction lengths of shielding upstream of the counters to 
filter out hadrons. 
Data from $D^+\rightarrow \overline{K}^{*0} \mu^{+}\nu _{\!_{\mu}}$ 
decays \cite{Chong} were used to choose selection criteria for muon 
candidates. 
Timing information from the smaller set of muon scintillation counters 
was used to improve the horizontal position resolution. Counter 
efficiencies, measured using muons originating from the primary target, 
were found to be $(99\pm 1)\%$ for the smaller counters and 
$(69\pm 3)\%$ for the larger counters. The probability for 
misidentifying a pion as a muon decreased with momentum, from about 
6$\%$ at 8 GeV/$c$ to $(1.3 \pm 0.1)\%$ above 20 GeV/$c$.

Events with evidence of well-separated 
production (primary) and decay (secondary) vertices were selected to 
separate charm candidates from background. Secondary vertices were 
required to be separated from the primary vertex by greater than 
$12\,\sigma_{_{\!L}}$, where $\sigma_{_{\!L}}$ 
is the calculated resolution of the measured longitudinal separation. 
Also, the secondary vertex had to be separated from the closest 
material in the target foils by greater than 
$5\,\sigma_{_{\!L}}^{\prime }$, where $\sigma_{_{\!L}}^{\prime }$ is 
the uncertainty in this separation. The vector sum of the momenta 
from secondary vertex tracks was required to pass within 
$40~\mu$m of the primary vertex in the plane perpendicular to the beam. 
Finally, the net momentum of the charm candidate transverse to 
the line connecting the production and decay vertices had to be 
less than 300 MeV/$c$. Decay track candidates were required to pass 
approximately 10 times closer to the secondary vertex than to the 
primary vertex. These selection criteria and kaon identification 
requirements were the same for both the search mode and for its 
normalization signal (discussed below). The mass ranges used for the 
resonant masses were: 
$\left| m_{\pi ^+\pi ^-} - m_{\rho ^0}\right| <150$ MeV/$c^2$, 
$\left| m_{K^-\pi ^+} - m_{\Kstar}\right| <55$ MeV/$c^2$, and 
$\left| m_{K^+K^-} - m_{\phi}\right| <10$ MeV/$c^2$.

To determine our selection cuts we used a ``blind'' analysis technique. 
Before the selection criteria were finalized, all events having masses 
within a window $\Delta M_S$ around the mass of the $D^{0}$ were 
``masked'' so that the presence or absence of any potential signal 
candidates would not bias our choice of selection criteria. All criteria 
were then chosen by studying events generated by a Monte Carlo (MC) 
simulation program and background events, outside the signal 
windows, from real data. The criteria were chosen to maximize the ratio 
$N_{MC}/\sqrt{N_B}$, where $N_{MC}$ and $N_B$ are the numbers of MC and 
background events, respectively, after all selection criteria were 
applied. The data within the signal windows were unmasked only after 
this optimization. We used asymmetric windows for the decay modes 
containing electrons to allow for the bremsstrahlung low-energy tail. 
The signal windows were: 
$1.83<M(D^0)<1.90$ GeV/$c^2$ for $\mu \mu $ and 
$1.76<M(D^0)<1.90$ GeV/$c^2$ for $ee$ and $\mu e$ modes.

We normalize the sensitivity of our search to topologically similar 
hadronic 3-body (resonant) or 4-body (non-resonant) decays. One 
exception to this is the case of 
$D^{0}\rightarrow \rho^{0} \ell ^{\pm }\ell ^{\mp }$ where we 
normalize to nonresonant  
$D^{0}\rightarrow \pi ^+\pi ^-\pi ^+\pi ^-$ because there is no published 
branching fraction for $D^0\rightarrow \rho ^{0}\pi ^+\pi ^-$. Table 
\ref{Norm} lists the normalization mode used for each signal mode and 
the fitted number of data events ($N_{\mathrm{Norm}}$).

The upper limit for each branching fraction $B_{X}$ is calculated using 
the following formulae:
\begin{equation}
B_{X}=\frac{N_{X}}{N_{\mathrm{Norm}}}
\frac{\varepsilon _{\mathrm{Norm}}}{\varepsilon _{X}}
\times B_{\mathrm{Norm}};~~ 
\frac{\varepsilon _{\mathrm{Norm}}}{\varepsilon _{X}}=
\frac{f_{\mathrm{Norm}}^{\mathrm{MC}}}{f_{X}^{\mathrm{MC}}}.
\label{BReqn}
\end{equation}
$N_{X}$ is the 90$\%$ confidence level (CL) upper limit on the 
number of decays for the rare or forbidden decay mode $X$ 
and $B_{\mathrm{Norm}}$ is the normalization mode branching 
fraction obtained from the Particle Data Group \cite{PDG}. 
$\varepsilon_{\mathrm{Norm}}$ and $\varepsilon_{X}$ are the detection 
efficiencies while  
$f_{\mathrm{Norm}}^{\mathrm{MC}}$ and $f_{X}^{\mathrm{MC}}$ are 
the fractions of Monte Carlo events that are reconstructed and pass 
the final selection criteria, for the normalization and decay modes 
respectively. 
\noindent
\parbox{0.48\textwidth}{
\begin{table}[b!]
\caption[]{
Normalization modes used.}
\label{Norm}
\vskip 5pt
\tabcolsep=4.0pt
\begin{tabular}{lll} 
Decay Mode&Normalization Mode&$N_{\mathrm{Norm}}$ \\
\hline
\vspace*{-8pt} &    & \\
$D^{0}\rightarrow \rho^{0} \ell ^{\pm }\ell ^{\mp }$&
$D^{0}\rightarrow \pi ^+\pi ^-\pi ^+\pi ^-$& 2049$\pm$53\\
$D^{0}\rightarrow \Kstar \ell ^{\pm }\ell ^{\mp }$&
$D^{0}\rightarrow \Kstar \pi ^+\pi ^-$& 5451$\pm$72\\ 
$D^{0}\rightarrow \phi \ell ^{\pm }\ell ^{\mp }$&
$D^{0}\rightarrow \phi \pi ^+\pi ^-$& 113$\pm$19\\ 
$D^{0}\rightarrow \pi \pi \ell\ell $& 
$D^{0}\rightarrow \pi ^+\pi ^-\pi ^+\pi ^-$& 2049$\pm$53\\
$D^{0}\rightarrow K\pi \ell\ell $&
$D^{0}\rightarrow K^-\pi ^+\pi ^-\pi ^+$&11550$\pm$113\\
$D^{0}\rightarrow KK\ell\ell $&
$D^{0}\rightarrow K^+K^-\pi ^+\pi ^-$& 406$\pm$41\\
\end{tabular}
\end{table}
}

The MC simulations use \pythjet~\cite{MC} as the physics 
generator and model the effects of resolution, detector geometry, 
magnetic fields, multiple scattering, interactions in the detector 
material, detector efficiencies, and the analysis selection criteria. 
The efficiencies for the normalization modes varied from approximately 
$0.2\%$ to $1\%$ depending on the mode, and the efficiencies for the 
search modes varied from approximately $0.05\%$ to $0.34\%$. We take 
muon and electron ID efficiencies from data.

Monte Carlo studies show that the experiment's acceptances are nearly 
uniform across the Dalitz plots, except that the dilepton 
identification efficiencies typically drop to near zero at the 
dilepton mass threshold. While the loss in efficiency varies channel by 
channel, the efficiency typically reaches its full value at masses 
only a few hundred MeV/$c^{\,2}$ above the dilepton mass threshold. We 
use a constant weak-decay matrix element when calculating the overall 
detection efficiencies. 

\begin{figure}
\vspace*{-18pt}
\centerline{\epsfxsize 3.375 truein \epsfbox{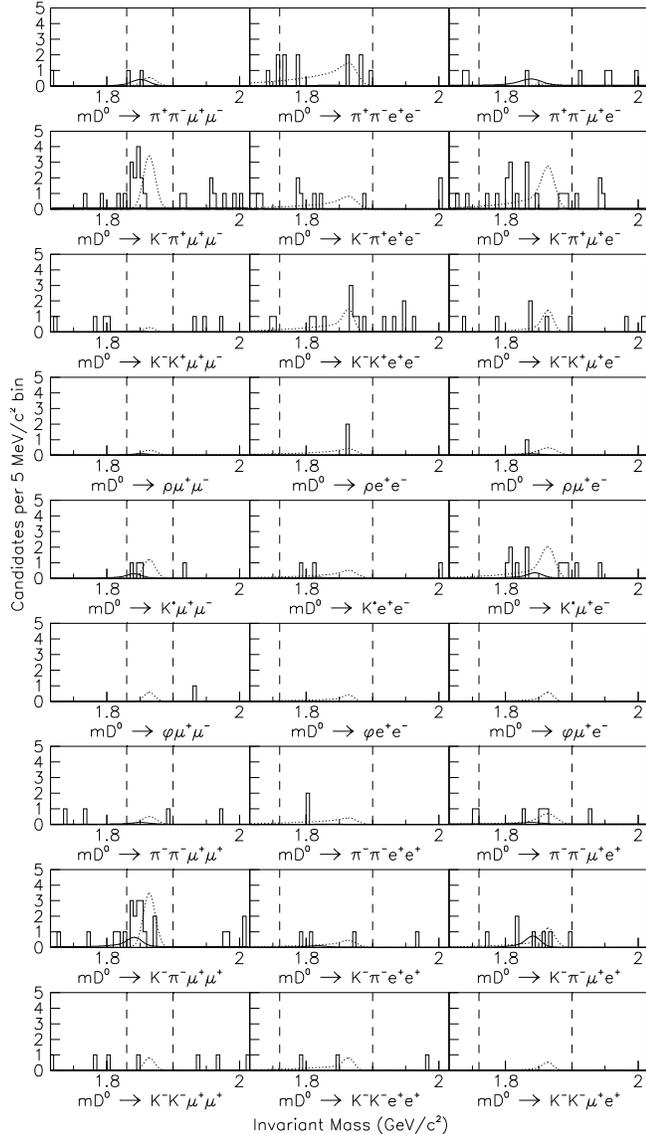}}
\caption[]{
\small Final event samples for the opposite signed dilepton (rows 1--3), 
resonant (rows 4--6), and same signed dilepton modes (rows 7--9) of 
$D^0$ decays. The solid curves display total estimated background; the 
dotted curves display signal shape for a number of events equal to 
the 90$\%$ CL upper limit. The dashed vertical lines are the 
$\Delta M_S$ boundaries.}
\label{Data}
\end{figure}

The 90$\%$ CL upper limits $N_{X}$ are calculated using the method of 
Feldman and Cousins \cite{Cousins} to account for background, and then 
corrected for systematic errors by the method of Cousins and Highland 
\cite{Cousins}. In these methods, the numbers of signal events are 
determined by simple counting, not by a fit. All results are shown in 
Fig.\ \ref{Data} and listed in Table \ref{Results}. Upper limits 
are determined using the number of candidate events observed and 
expected number of background events within the signal region.  

Background sources that are not removed by the selection criteria 
discussed earlier include decays in which hadrons (from real, 
fully-hadronic decay vertices) are misidentified as leptons. These 
misidentified leptons can come from hadronic showers reaching muon 
counters, decays-in-flight, and random overlaps of tracks from 
otherwise separate decays (``accidental'' sources). In the case where 
kaons are misidentified as pions or leptons, candidate masses shift 
below signal windows. However, we remove these events to prevent 
them from influencing our background estimate, which is 
partially obtained from the mass sidebands (see discussion of 
$N_{\mathrm{Cmb}}$ below). To remove these events prior to the 
selection-criteria optimization, we reconstruct all candidates as each 
of the non-resonant normalization modes and test whether the 
masses are consistent with $m_{D^{0}}$. If so, we remove the events, 
but only if the number of kaons in the final state differs from that 
of the search mode. We do not remove events having the same number of 
kaons, as the loss in acceptance for true signal events would be 
excessive.

There remain two sources of background: hadronic decays where pions are 
misidentified as leptons ($N_{\mathrm{MisID}}$) and ``combinatoric'' 
background ($N_{\mathrm{Cmb}}$) arising primarily from false vertices 
and partially reconstructed charm decays. The background 
$N_{\mathrm{MisID}}$ arises from the normalization modes. To estimate 
the rate for misidentifying $\pi \pi $ as $\ell \ell $, for all but the 
$D^0\rightarrow K^-\pi ^+\ell ^+\ell ^-$ modes, we assume all 
$D^0\rightarrow K^-\pi ^+\ell ^+\ell ^-$ candidates observed (after 
subtracting combinatoric background estimated from mass sidebands) 
result from misidentification of $D^0\rightarrow K^-\pi ^+\pi ^-\pi ^+$ 
decays and count the number of $D^0\rightarrow K^-\pi ^+\ell ^+\ell ^-$ 
decays passing the final selection criteria. We then divide by twice 
the number of $D^0\rightarrow K^-\pi ^+\pi ^-\pi ^+$ normalization 
events with $K^-\pi ^+\ell ^+\ell ^-$ mass within $\Delta M_S$ 
boundaries (twice because there are two possible $\pi ^+$ 
misidentifications). 

From this procedure, the following misidentification rates were 
obtained: 
$r_{\mu\mu}= (3.4 \pm 2.4)\times 10^{-4}$, 
$r_{\mu e}= (4.2 \pm 1.4)\times 10^{-4}$, and 
$r_{ee}= (9.0 \pm 6.2 )\times 10^{-5}$. 
For modes in which two possible pion combinations can contribute, \eg, 
$D^0\rightarrow K^-\pi ^{+}\mu ^{\pm}\mu ^{\mp}$, we use twice the 
above rate; and for $D^{0}\rightarrow \pi ^+\pi ^-\pi ^+\pi ^-$, where 
there are 4 possible combinations, we use 4 times this rate in 
calculating $D^{0}\rightarrow \pi ^+\pi ^-\ell ^+\ell ^-$. 
Using these rates, we estimate the numbers of misidentified candidates, 
$N_{\mathrm{MisID}}^{V\ell\ell}$ and $N_{\mathrm{MisID}}^{hh\ell\ell}$, 
in the signal windows as follows:
\vspace*{-3pt}
\begin{equation}
N_{\mathrm{MisID}}^{hh\ell\ell} =  r_{\! _{\ell\ell}} 
\times N_{\mathrm{Norm}}^{hh\pi\pi} 
{\rm \, and\, } N_{\mathrm{MisID}}^{V\ell\ell} = r_{\! _{\ell\ell}} 
\! \times \! N_{\mathrm{Norm}}^{V\pi\pi}, 
\label{Nmidid}
\end{equation}
where $N_{\mathrm{Norm}}^{hh\pi\pi}$ and $N_{\mathrm{Norm}}^{V\pi\pi}$ 
are the numbers of normalization hadronic decay candidates in the 
signal windows. 

To calculate the upper limits for the 
$D^0\rightarrow K^-\pi ^+\ell ^+\ell ^-$ modes, we set 
$N_{\mathrm{MisID}}$ to zero as we do not have an independent estimate 
of the misidentification rates. This results in conservative upper 
limits. If we had used the misidentification rates from our previous, 
3-body decay study \cite{FCNCnew}, then our limits for the three 
$D^0\rightarrow K^-\pi ^+\ell ^+\ell ^-$ modes would be lower by about 
a factor of two. 

To estimate the combinatoric background $N_{\mathrm{Cmb}}$ within a 
signal window $\Delta M_S$, we count events having masses within an 
adjacent background mass window $\Delta M_B$, and scale this number 
($N_{\Delta M_B}$) by the relative sizes of these windows:
$N_{\mathrm{Cmb}} = ({\Delta M_S}/{\Delta M_B}) \times N_{\Delta M_B}$. 
To be conservative in calculating our 90$\%$ confidence level upper 
limits, we take combinatoric backgrounds to be zero when no 
events are located above the mass windows. Table \ref{Results} shows 
the numbers of combinatoric background, misidentification 
background, and observed events for all 27 modes.
\noindent
\parbox{0.48\textwidth}{
\begin{table}
\vspace*{-6pt}
\caption[]{
E791 90$\%$ confidence level (CL) upper limits on the number of events 
and branching fraction limits ($\!\times \!10^{-5}$). Previously 
published limits \cite{PDG,CLEO} for the nine $D^0 \to V\ell^+\ell^-$ 
modes are 23, 10, 4.9, 118, 14, 10, 41, 5.2, and 3.4 
$\times \, 10^{-5}$.
}
\label{Results}
\vskip 1pt
\tabcolsep=2.0pt
\begin{tabular}{lcccccr}
Mode & \multicolumn{2}{c}{(Est.~BG)} & 
&Sys.&&\small{E791}\\
$D^{0}\rightarrow $&\small{$N_{\mathrm{Cmb}}$}&
\small{$N_{\mathrm{MisID}}$}&\small{$N_{\mathrm{Obs}}$}&
Err.&\small{$N_{X}$}&\small{Limit}\\
\hline
\vspace*{-10pt} &    &   &     &     &    &     \\
 $\pi ^{+}\pi ^{-}\mu ^{+}\mu ^{-}$&0.00&3.16&2&11$\%$
 &2.96&$3.0$\\
 $\pi ^{+}\pi ^{-}e^{+}e^{-}$&0.00&0.73&9&12$\%$
 &15.2&$37.3$\\
 $\pi ^{+}\pi ^{-}\mu ^{\pm }e^{\mp }$&5.25&3.46&1&15$\%$
 &1.06&$1.5$\\
 $K^{-}\pi ^{+}\mu ^{+}\mu ^{-}$&3.65&0.00&12&11$\%$
 &15.4&$35.9$\\
 $K^{-}\pi ^{+}e^{+}e^{-}$&3.50&0.00&6&15$\%$
 &7.53&$38.5$\\
 $K^{-}\pi ^{+}\mu ^{\pm }e^{\mp }$&5.25&0.00&15&12$\%$
 &17.3&$55.3$\\
 $K^{+}K^{-}\mu ^{+}\mu ^{-}$&2.13&0.17&0&17$\%$
 &1.22&$3.3$\\
 $K^{+}K^{-}e^{+}e^{-}$&6.13&0.04&9&18$\%$
 &9.61&$31.5$\\
 $K^{+}K^{-}\mu ^{\pm }e^{\mp }$&3.50&0.17&5&17$\%$
 &6.61&$18$\\
\hline
\vspace*{-10pt} &    &   &     &     &    &\\
 $\rho ^{0}\mu ^{+}\mu ^{-}$&0.00&0.75&0&10$\%$
 &1.80&$2.2$\\
 $\rho ^{0}e^{+}e^{-}$&0.00&0.18&1&12$\%$
 &4.28&$12.4$\\
 $\rho ^{0}\mu ^{\pm }e^{\mp }$&0.00&0.82&1&11$\%$
 &3.60&$6.6$\\
 $\Kstar \mu ^{+}\mu ^{-}$&0.30&1.87&3&24$\%$
 &5.40&$2.4$\\
 $\Kstar e^{+}e^{-}$&0.88&0.49&2&25$\%$
 &4.68&$4.7$\\
 $\Kstar \mu ^{\pm }e^{\mp }$&1.75&2.30&9&24$\%$
 &12.8&$8.3$\\
 $\phi \mu ^{+}\mu ^{-}$&0.30&0.04&0&33$\%$
 &2.33&$3.1$\\
 $\phi e^{+}e^{-}$&0.00&0.01&0&33$\%$
 &2.75&$5.9$\\
 $\phi \mu ^{\pm }e^{\mp }$&0.00&0.05&0&33$\%$
 &2.71&$4.7$\\
\hline
\vspace*{-10pt} &    &   &     &     &    &\\
 $\pi ^{-}\pi ^{-}\mu ^{+}\mu ^{+}$&0.91&0.79&1&9$\%$
 &2.78&$2.9$\\
 $\pi ^{-}\pi ^{-}e^{+}e^{+}$&0.00&0.18&1&11$\%$
 &4.26&$11.2$\\
 $\pi ^{-}\pi ^{-}\mu ^{+}e^{+}$&2.63&0.86&4&10$\%$
 &5.18&$7.9$\\
 $K^{-}\pi ^{-}\mu ^{+}\mu ^{+}$&2.74&3.96&14&9$\%$
 &15.7&$39.0$\\
 $K^{-}\pi ^{-}e^{+}e^{+}$&0.88&1.04&2&16$\%$
 &4.14&$20.6$\\
 $K^{-}\pi ^{-}\mu ^{+}e^{+}$&0.00&4.88&7&11$\%$
 &7.81&$21.8$\\
 $K^{-}K^{-}\mu ^{+}\mu ^{+}$&1.22&0.00&1&17$\%$
 &3.27&$9.4$\\
 $K^{-}K^{-}e^{+}e^{+}$&0.88&0.00&2&17$\%$
 &5.28&$15.2$\\
 $K^{-}K^{-}\mu ^{+}e^{+}$&0.00&0.00&0&17$\%$
 &2.52&$5.7$\\
\end{tabular}
\end{table}
}

The sources of systematic errors in this analysis include: errors from 
the fit to the normalization sample $N_{\mathrm{Norm}}$; statistical 
uncertainty on the selection efficiencies, calculated for Monte Carlo 
simulated events, for both 
$f_{\mathrm{Norm}}^{\mathrm{MC}}$ and $f_{X}^{\mathrm{MC}}$; 
uncertainties in the calculation of misidentification background; and 
uncertainties in the relative efficiency for each mode, including lepton 
tagging efficiencies. These tagging efficiency uncertainties include: 1) 
muon counter efficiencies from hardware performance;  and 2) the 
fraction of signal events (based on simulations) that would remain 
outside the signal window due to bremsstrahlung tails. Also, for the 
$D^0\rightarrow \rho ^{0}\ell ^+\ell ^+$ modes, an additional systematic 
error is included because we are using 
$D^{0}\rightarrow \pi ^+\pi ^-\pi ^+\pi ^-$ as the normalization mode 
since there is no published branching fraction for 
$D^0\rightarrow \rho ^{0}\pi ^+\pi ^-$. The sums, taken in quadrature, 
of these systematic errors are listed in Table \ref{Results}.

In summary, we use a ``blind'' analysis of data from Fermilab 
experiment E791 to obtain upper limits on the dilepton branching 
fractions for 27 flavor-changing neutral current, lepton-number 
violating, and lepton-family violating decays of $D^0$ mesons. No 
evidence for any of these 3 and 4-body decays is found. Therefore, we 
present upper limits on the branching fractions at the 90$\%$ 
confidence level. Four limits represent significant improvements over 
previously published results. Eighteen of these modes have no previously 
reported limits.

We thank the staffs of Fermilab and 
participating institutions. This research was supported by 
the Brazilian Conselho Nacional de Desenvolvimento Cient\'\i fico e 
Tecnol\'ogico, CONACyT (Mexico), the Israeli Academy of Sciences and 
Humanities, the U.S.-Israel Binational 
Science Foundation, and the U.S.~National Science Foundation 
and Dept.~of Energy. 
The Universities Research Assn.~operates Fermilab for 
the U.S.~Dept.~of Energy.

\bibliographystyle{unsrt}

\end{multicols}
\end{document}